# Low temperature annealing method for fabricating alloy nanostructures and metasurfaces: Unlocking a novel degree of freedom

Debdatta Ray[†], Hsiang-Chu Wang[†], Jeonghyeon Kim, Christian Santschi, Olivier J.F Martin[*,]

Nanophotonics and Metrology Laboratory, Swiss Federal Institute of Technology, Lausanne (EPFL), 1015 Lausanne, Switzerland

[†]These authors contributed equally to this work

[*]corresponding author: olivier.martin@epfl.ch

## Abstract:

The material and exact shape of a nanostructure determine its optical response, which is especially strong for plasmonic metals. Unfortunately, only a very few plasmonic metals are available, which limits the spectral range where these strong optical effects can be utilized. Alloying different plasmonic metals can overcome this limitation, at the expense of using a high temperature alloying process, which adversely destroys the nanostructure shape. Here, we develop a low temperature alloying process at only 300°C and fabricate Au-Ag nanostructures with a broad diversity of shapes, aspect ratios and stoichiometries. EDX and XPS analyses confirm the homogeneous alloying through the entire sample. Varying the alloy stoichiometry tunes the optical response of the nanostructure and controls spectral features such as Fano resonances. Binary metasurfaces that combine nanostructures with different stoichiometries are fabricated using multiple-step electron beam lithography, and their optical function as hologram or Fresnel zone plate is demonstrated at the visible wavelength of $\lambda = 532$ nm. This low temperature annealing technique provides a versatile and cost-effective way of fabricating complex Au-Ag nanostructures with arbitrary stoichiometry.

## Introduction

Every nanotechnology is indissociable from the materials it uses. Conversely, the materials involved often determine and limit the possible technological choices and the potential applications that can be envisioned. This is true for many different scientific fields, from energy[1,2] to chemistry[3,4], semiconductors[5-7], medicine[8,9] or even food industry[10]. Plasmonics – the resonant optical interaction of light with free electrons in metals – also strongly relies on the utilization of specific materials, viz. coinage metals with a very high free electron density[11].

Thanks to its electronic properties, ease of fabrication and chemical stability, Au is the most widely used coinage metal for plasmonic applications. Its dielectric function $\varepsilon(\lambda)$, where $\lambda$ is the wavelength of light, produces a plasmon resonance in the long-wavelength range of the visible spectrum. Other metals, like Al and Ag, have plasmon resonances towards shorter wavelengths, but are more difficult for nanotechnology[12]. While the real part of $\varepsilon(\lambda)$ determines the wavelength at which the plasmon resonance occurs, its imaginary part controls the plasmon resonance strength[13]. Since a decade, there has been a surge of research for alternative materials to Au, Ag and Al, in order to exploit plasmon resonances over the entire visible and near infrared spectrum[14-16].

In addition to pure materials, heterogeneous combinations of metals such as core-shell particles[17] and alloys[18] have also been proposed. Especially Au-Ag alloys have demonstrated extremely interesting properties in terms of dielectric function tunability and chemical stability[19,20]. The capability of alloys to adjust the optical properties of the individual metals by altering their electronic band structure[18,21-23], is the driving force behind a large number of applications, including SERS[19,20,24-26], biosensing[27-30], therapeutics[21], photochemistry[31-33], colour generation[34] and generation of hot electrons[35]. Moreover, Au-Ag alloys are chemically more stable than pure Ag[19] and Au-Ag alloy very well thanks to their similar lattice constants[36] and a very simple binary phase diagram[37].

Existing techniques for the fabrication of Au-Ag alloys include laser ablation[38], co-reduction of Au-Ag salts[39,40], co-deposition[41,42], annealing at very high temperatures of Au-Ag core-shell nanoparticles with[43-45] and without[46] $SiO_2$ coating. All those bottom-up techniques provide limited control of the nanoparticles shape, as well as its reproducibility. In contrast, top-down fabrication techniques, such as e-beam lithography (EBL), are remarkably versatile and provide reliable control over the shape and size of individual nanoparticles[47]. Since the size of targeted nanostructures is typically in the range of a few tens to hundreds of nm, metal depositions are carried out using directional e-beam evaporation rather than isotropic co-sputtering to ensure reliable lift-off performances. Alloys have been deposited either by alloyed targets[48], which is costly since each composition requires a different target; or by alternating deposition of thin Au and Ag layers with subsequent annealing at high temperature, resulting in nanostructures that do not retain the original shape[49]. This represents a very severe limitation since the shape of plasmonic nanostructures determines their optical properties[13]. This warrants a need for cost-effective techniques for the stable fabrication of alloyed nanostructures with well-controlled sizes and shapes.

Here, we address this challenge and propose an original process for fabricating alloyed nanostructures. The method is based on the deposition of a bilayer of Au and Ag with subsequent annealing at low temperature at 300°C, far below the melting points of the individual metals. Thanks to the low temperature, the shape of the nanoparticles is retained after annealing. To demonstrate and verify that homogeneous alloys can be fabricated by this low temperature process, we first fabricate thin films of Au-Ag alloys and confirm their homogeneity by X-ray photoelectron spectroscopy (XPS) and energy-dispersive X-ray spectroscopy (EDX). Ellipsometric studies demonstrate the variation of the permittivity across different compositions. This technique is then applied to fabricate different alloyed nanostructures that perfectly retain their shape after low temperature annealing. Optical measurements demonstrate that the dielectric function of these nanostructures is tuned by varying the alloy's composition. Finally, metasurfaces that combined different alloys are realized and their optical function as hologram or Fresnel zone plate is demonstrated experimentally in the visible part of the spectrum.

## Alloyed thin films

Au-Ag nanostructures are normally alloyed at temperatures in the order of 800-900 °C, close to the melting point of the metals[22,44,50]. To demonstrate as a first step the successful annealing at a temperature as low as 300°C, thin Au-Ag films are fabricated, as described in Fig. 1a. The total film thickness is 150 nm and is kept constant for the different compositions. The latter are determined by depositing different thicknesses of Au and Ag using the e-beam evaporator LAB600H (Leybold-Optics).

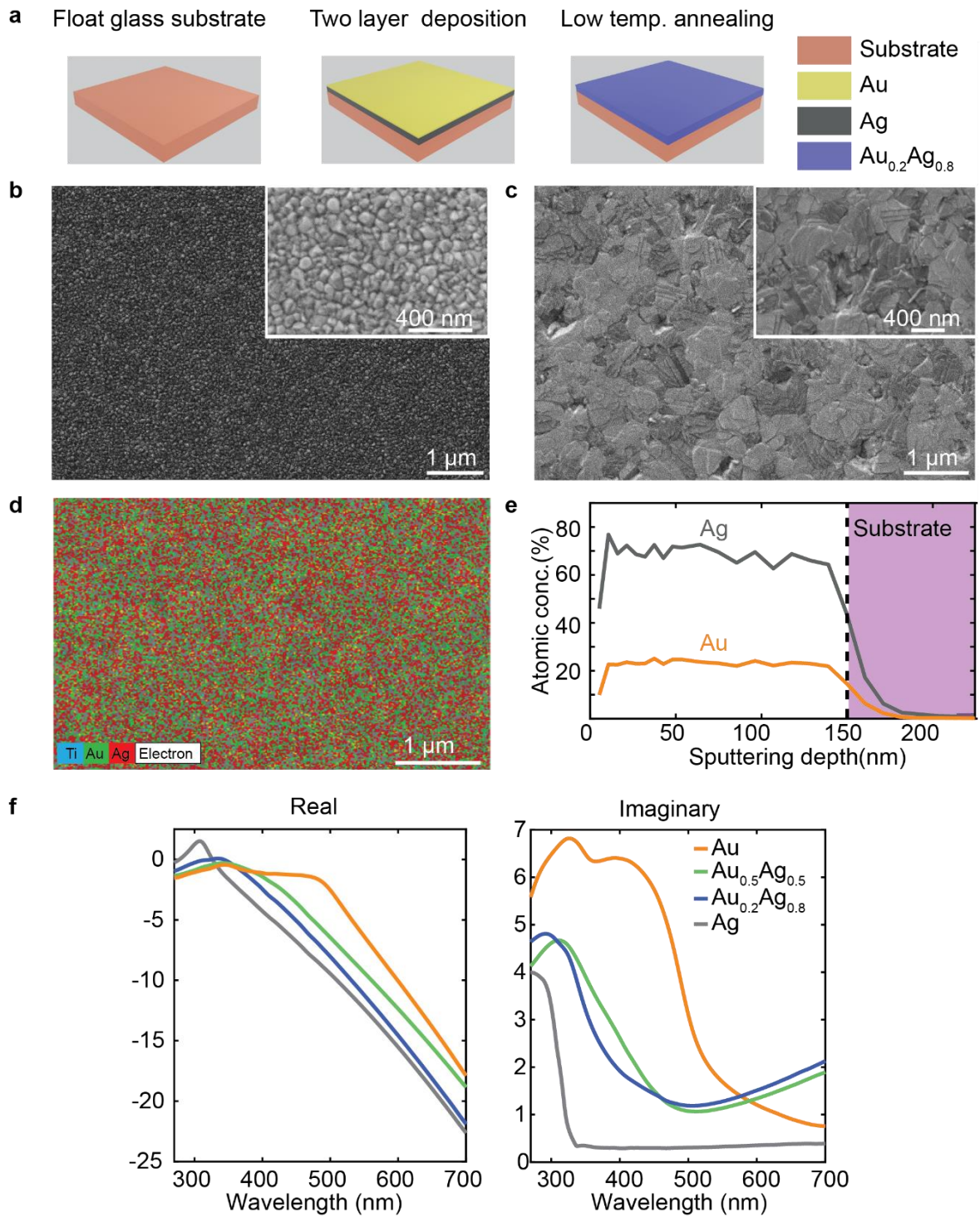

**Fig. 1 | Demonstration of low temperature alloyed thin films. a** Process flow for fabricating thin films with a specific stoichiometry determined by the thicknesses of the different metal layers. **b** SEM image of a double-layer film for $Au_{0.2}Ag_{0.8}$ before annealing. **c** SEM image of the same film after annealing. The insets in **b** and **c** show a magnified view. **d** Superimposed EDX maps for Au, Ag and Ti. **e** XPS analysis of the $Au_{0.2}Ag_{0.8}$ alloy,

demonstrating a homogeneous alloy over the entire 150 nm depth. **f** Experimentally obtained ellipsometric data for pure Au, $Au_{0.2}Ag_{0.8}$, $Au_{0.5}Ag_{0.5}$ and pure Ag.

To illustrate this process, let us consider the fabrication of an $Au_{0.2}Ag_{0.8}$ alloy: 120 nm of Ag and 30 nm of Au are deposited on a float glass substrate coated with 2 nm of Ti acting as an adhesion layer. Immediately afterwards, the entire film is annealed under $N_2$ atmosphere in an oven (Neytech Qex Vacuum Porcelain furnace) at 300 °C for 8 hours followed by 30 min at 450 °C. Trial and error has demonstrated that this additional 30 min annealing at 450 °C homogenizes the alloy: XPS results for a 150 nm $Au_{0.2}Ag_{0.8}$ film annealed only at 300 °C for 8 hours without this additional step indicate that the film is inhomogeneous, Supplementary Fig. SI 1a. Figures 1b and 1c show SEM images for this double layer film before and after annealing (300 °C for 8 hours followed by 30 min at 450 °C), with insets providing magnified views. It can be clearly seen that the film has lost its grainy appearance and reorganized into a more crystalline configuration during the annealing process[51,52]. EDX maps of all elements of interest are superimposed in Fig. 1d, revealing a homogeneous distribution of Au and Ag. XPS analysis demonstrates a homogeneous alloy over the entire 150 nm depth, Fig. 1e. In contrast, XPS analysis of films carried out prior to annealing distinctly shows the bilayer of Ag and Au, Supplementary Fig. SI. 1b. Thus, XPS and EDX results confirm that we have a homogeneous Au-Ag distribution after annealing the samples at low temperature, even for a relatively thick film. Figure 1f reports the experimentally obtained ellipsometric data for Au, $Au_{0.2}Ag_{0.8}$, $Au_{0.5}Ag_{0.5}$ and Ag. It can be seen that the real part of the dielectric function for both alloys lies between pure Au and pure Ag. This modification of $\varepsilon_{real}$ with change in composition leads to variations of the plasma frequency as well[53]. The imaginary part, however, is more complicated since it reflects the material losses and can be affected by various factors such as grain boundaries and defect sites[54,55]. Ellipsometry and numerical analysis were carried out using SOPRA GES 5E and WINELI II software, respectively. For all the measurements, the incident angle was kept at 75° and both real and imaginary parts of the dielectric function were extracted from the amplitude component $\Phi$ and the phase difference $\Delta$[12]. The experimental data are compared in Supplementary Fig. SI 2 with the permittivity model of Rioux *et al.*[56], which agrees very well. The experimental data agree also well with the experimental values reported by Rodriguez. *et al.* [57].

Before we proceed to more results, we would like to briefly discuss the conditions that make alloying at low temperature successful. Homogeneously alloyed samples at 300 °C – which is well below the Tamman temperature of Au and Ag [58] – require the annealing to be performed within a few days after metal deposition. The deposition of the metal film, leaves it in an unstable morphology due to the presence of high relative humidity and the film remains in this state for a couple of days after the deposition[51]. Since the film is not yet well-ordered, it takes less thermal energy for the Au and Ag atoms to diffuse. Thus, the onset of the diffusion occurs at lower temperatures, which makes it possible to completely alloy thin films at a temperature of 300 °C. On the other hand, we have observed that it was not possible to anneal at low temperature bilayer Au-Ag films that were kept in a nitrogen atmosphere (Saito Keiryoki MFG. CO. LTD) for more than a week before annealing. This is shown in Supplementary Fig. SI 3, with SEM images before and after annealing for 150 nm $Au_{0.8}Ag_{0.2}$ thin films kept in nitrogen atmosphere for 4.5 months after deposition. After annealing, nanoparticles are formed on the film (Supplementary Fig. SI 3d), as opposed to a homogeneous alloyed film that was obtained when annealed within a few days of deposition (Supplementary Fig. SI 3b). Moreover, the appearance of the not-annealed bilayer film after 4.5 months is very different (more smooth and flat, Supplementary Fig. SI 3c) compared to that after only a few days (more grainy, Supplementary Fig. SI 3a). This indicates that with passing time the bilayer film becomes very stable and thus requires a higher temperature to form an homogeneous alloy.

## Alloyed nanostructures

The advantages associated with the low temperature annealing presented in this article are unleashed for the fabrication of nanostructures with different stoichiometries, as illustrated in Fig. 2a. Similar to the alloyed thin films the Au-Ag bilayers are deposited by e-beam evaporation. Figures 2b and 2c show SEM images of an array of 50 nm thick rods and triangles before and after annealing. The corresponding insets magnify a single 400 nm x 50 nm nanorod and a nanotriangle with a 260 nm side length. It is remarkable that the nanostructure shapes are perfectly retained after annealing, even for very high aspect ratio structures. The yield of this approach is also very high, as indicated by the widefield images. EDX measurements shown in Supplementary Fig. SI 4 reveal a homogeneous Au-Ag distribution over the entire nanostructure. The benefits of the low temperature annealing process can be better appreciated by looking at the SEM images for similar nanostructures obtained by high temperature annealing at 800°C for 3 hours in a nitrogen environment (bottom row of Fig. 2b and 2c). Those nanostructures do not retain their shape at all, as opposed to those of low temperature annealing. Consequently, nanostructures with a specific shape, annealed at high temperature will not exhibit the optical response they were designed for[59].

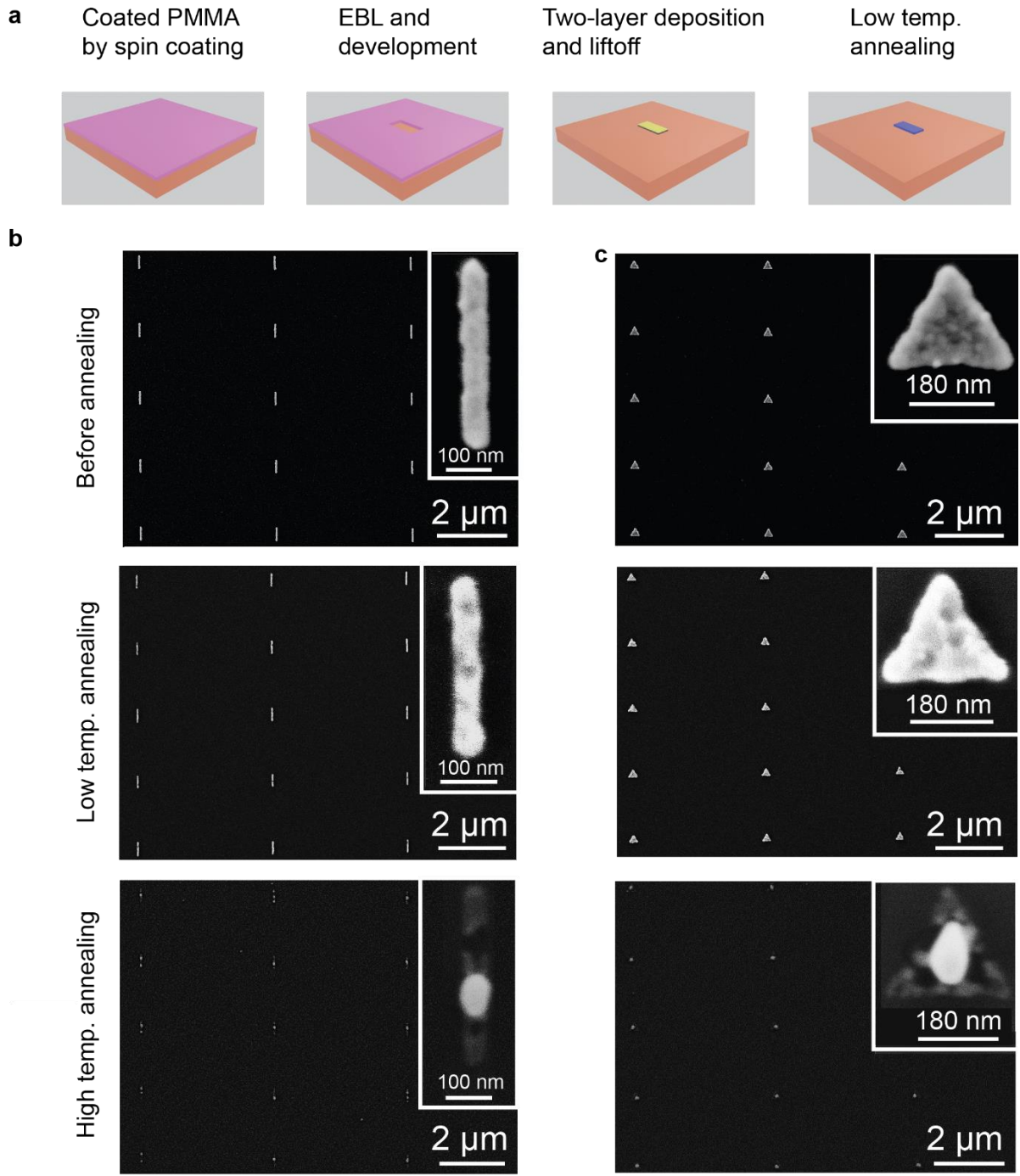

**Fig. 2 | Fabrication and optical characterization of alloyed nanostructures. a** Process flow (same colors as in Fig. 1a). SEM images of an array of $Au_{0.2}Ag_{0.8}$ **b** 400 nm x 50 nm rods and **c** 260 nm equilateral triangles before (top row), after (middle row) low temp annealing at 300°C for 8 hours followed by 450°C for 30 min, as well as (bottom row) high temp. annealing at 800°C for 3 hours. The insets show a single nanostructure from the corresponding array.

More sophisticated arrangements like 4-rod Fano-resonant structures are fabricated[51,60], Fig. 3a. The lengths of the longer and shorter rods are 130 nm and 68 nm, respectively. The gaps are 30 nm and the structures thickness is 40 nm. After annealing, the shape is perfectly retained. The Fano spectrum arises from the interference of a broad bright and narrow dark mode[60] and can be tuned by changing the alloy composition. Exactly the same geometry was fabricated in 5 different stoichiometries: Au, $Au_{0.8}Ag_{0.2}$, $Au_{0.5}Ag_{0.5}$, $Au_{0.2}Ag_{0.8}$ and Ag. The optical responses of these structures measured in dark field microscope (setup described in Supplementary Fig. SI 5) are shown in Fig. 3c. The experiments confirm the variation of the spectra with the composition. Simulations were carried out using the periodic surface integral equation method[61,62] and the dielectric functions obtained from the Rioux model[56]. The agreement between simulations and experiments is fair with some spectral features washed out by inhomogeneous broadening associated with thefabrication error and imperfections, Fig. 3c. The results presented in this section demonstrate the additional degree of freedom offered by the fabrication of alloyed Au-Ag nanostructures with varying stoichiometry.

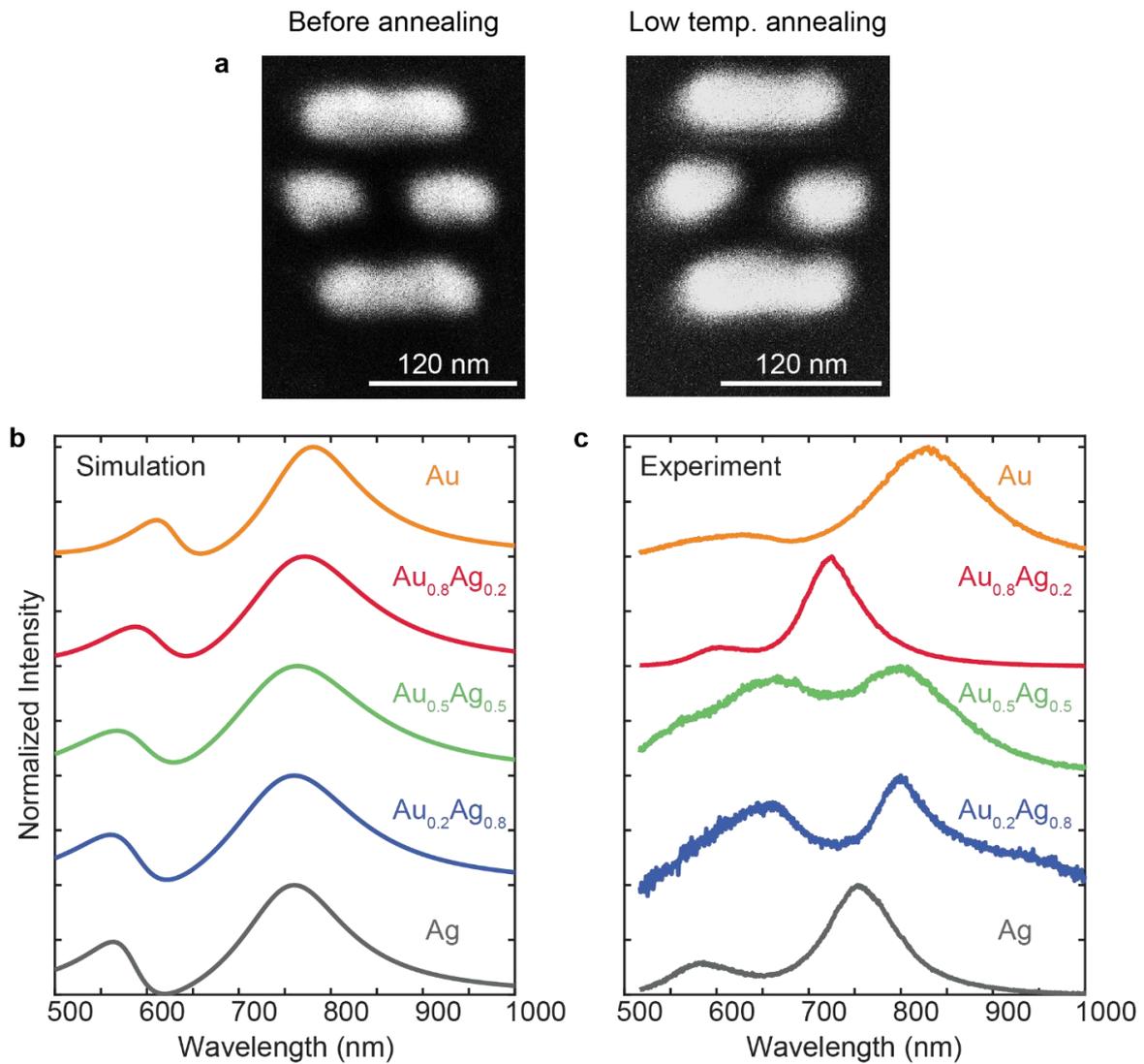

**Fig. 3 | Fabrication and optical characterization of Fano resonant structures. a** SEM images of a $Au_{0.8}Ag_{0.2}$ 4-rod Fano-resonant structure before and after low. temp annealing. **b** Calculated and **c** measured optical spectra of the Fano-resonant structures with different stoichiometries.

## Alloyed metasurfaces

Let us now apply this low temperature annealing technique to the fabrication of metasurfaces like holograms and Fresnel zone plates, which require combining alloyed nanostructures with different stoichiometries on the same substrate. Generally, tuning the response of a meta-atom is achieved by tailoring its geometry to produce the desired phase and amplitude response[63]. Here, instead of varying the shape, we change the material whilst keeping the shape constant. Binary phase metasurfaces are fabricated with two different alloys, $Au_{0.8}Ag_{0.2}$ and $Au_{0.2}Ag_{0.8}$. It should be noted that in this work the meta-atoms shape is kept constant to emphasize the effect of changing the optical properties of the material; this nanotechnology can of course be applied to meta-atoms with different shapes as well as different stoichiometries, thus providing a very large number of degrees of freedom for the metasurface design.

With Au-Ag alloys, it is straightforward to develop metasurfaces that operate in the visible; here we chose $\lambda$ = 532 nm and fabricate a periodic array of meta-atoms with periodicity p = 300 nm to avoid any diffraction order. Alloyed discs $Au_xAg_{1-x}$ with diameter d = 110 nm and thickness h = 50 nm, atop an Al mirror (thickness 150 nm) with a t = 100 nm thick $SiO_2$ spacer are used as meta-atoms, see the inset in Fig. 4a. Such a structure is polarization insensitive and supports a gap plasmon[64,65], which enables the re-radiation of the incident wave with a phase shift that can be controlled over a $2\pi$-range[66]. The reflectance and phase shift are computed as a function of the wavelength using the periodic surface integral equation method[61,62], Fig. 4a. The permittivity for Al and $SiO_2$ originate from the literature[67], whereas that for $Au_xAg_{1-x}$ are computed using Rioux's model[56]. Figure 4a indicates that at $\lambda$ = 532 nm the reflectance is the same for $Au_{0.2}Ag_{0.8}$ and $Au_{0.8}Ag_{0.2}$ nanostructures, while the phase they produce is shifted by 165°; this nanostructures pair can therefore serve as meta-atoms for a binary metasurface[68]. This is confirmed in Fig. 4b by measurements on alloyed disc arrays with five different stoichiometries (reflectance measurement setup described in Supplementary Fig. SI 6). The agreement between these measurements and simulations is good, especially the spectral locations of the reflection dip agree very well.

Having demonstrated the ability to manipulate the reflected light using arrays of alloyed nanostructures, let us now fabricate a meta-hologram and a Fresnel zone plate using the stoichiometries $Au_{0.8}Ag_{0.2}$ and $Au_{0.2}Ag_{0.8}$ to realize the two different disc-shaped meta-atoms. For the meta-hologram, we calculate the phase distribution using the computer generated hologram method[69]. Since the meta-hologram is built from two different alloys, a 3-step EBL process illustrated in Supplementary Fig. SI 7 is required to position the corresponding nanostructures with different compositions on the metasurface. In the first step, alignment marks are written and in the consecutive 2nd, respectively 3rd steps, nanostructures with $Au_{0.8}Ag_{0.2}$, respectively $Au_{0.2}Ag_{0.8}$, are deposited. The annealing is performed after the final step, once all the meta-atoms with different stoichiometries have been deposited on the surface. The SEM images of the hologram device in Fig. 4c indicate that all the unit cells share the same physical dimensions and the different meta-atoms are well aligned over a very large area (the inset highlights with colours the two different stoichiometries used in the metasurface). This is quite remarkable since these nanostructures are written in two successive, independent steps, including photoresist coating, EBL, development, two-layer metal deposition and lift-off. Images of the hologram projection under polarized laser illumination are shown in Figs. 4d and e for two different incident polarizations (setup described in Supplementary Fig. SI 8). The holograms are polarization-independent thanks to the symmetrical arrangement of the meta-atoms.

For the Fresnel zone plate, the spatial phase profile follows the equation

$$\varphi(r) = \frac{2\pi}{\lambda}\left(\sqrt{r^2 + f^2} - f\right), \quad (1)$$

where $r$ is the distance from the centre of the lens and $f$ represents the focal length[70]. For demonstration purposes, the lens is designed with a NA = 0.35 in air. The lens diameter is D = 300 µm and the focal length is $f$ = 400 µm. Figure 4f shows the SEM image of the fabricated lens. The inset in Fig. 4f provides a magnified view of the centre part of the lens, which is composed of meta-atoms with the same dimensions but different Au-Ag stoichiometries. The focal spot, displayed in Fig. 4g, is measured in the Fourier plane whereas the intensity profile, shown in Fig. 4h, is determined in the focal plane using the setup shown in Supplementary Fig. SI 9. The full width at half maximum (FWHM) is 1.8 µm, which agrees extremely well with the theoretical diffraction spot size $D_{\text{FWHM}} = 1.22\lambda/\text{NA} = 1.85\,\mu\text{m}$.

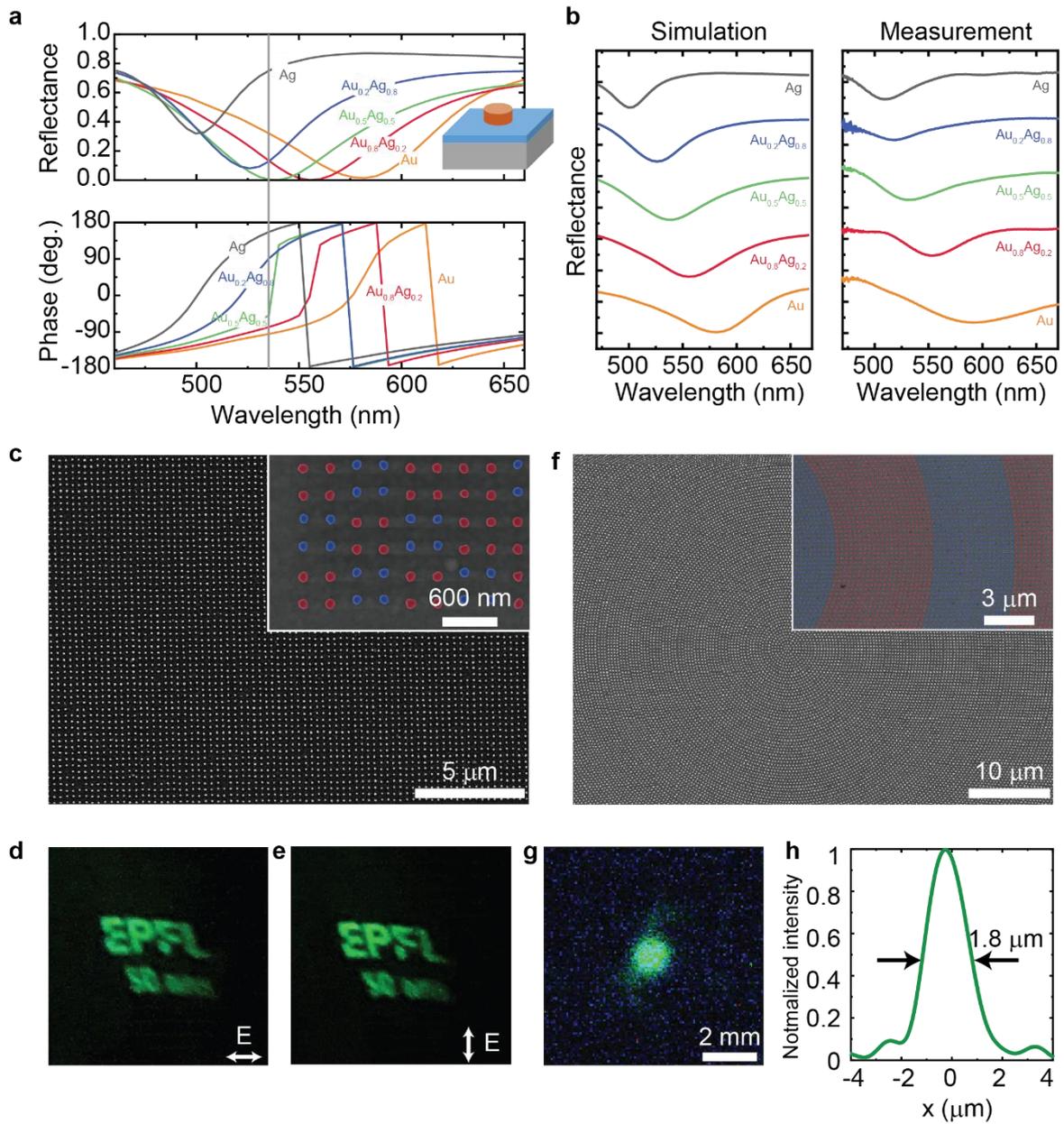

**Fig. 4 | Alloyed metasurfaces. a** Simulated reflectance and phase for alloyed meta-atoms with different stoichiometries. The inset shows the meta-atom composed of an alloyed disc with diameter d = 110 nm, height h = 50 nm, positioned atop an Al mirror with a $SiO_2$ spacer of thickness t = 100 nm. The period is p = 300 nm. **b** Simulated and measured reflectance spectra for different disc stoichiometries. **c** SEM image of the hologram device and its optical projection under **d** horizontally and **e** vertically polarized light. **f** SEM image of the FZP with focal length $f$ = 400 μm and NA = 0.35. **g** Intensity distribution at the focal plane and **h** normalized intensity profile at the centre of the focal spot. The insets in panels **c** and **f** emphasize with colours the two different stoichiometries used for the meta-atoms.

## Conclusions

We have introduced a facile and cost-effective nanofabrication technique to produce $Au_xAg_{1-x}$ alloyed nanostructures with arbitrary stoichiometry solely determined by the thickness of the different deposited metals. Thanks to the low temperature annealing process, well-alloyed nanostructures with arbitrary shape can be fabricated, including those with a high aspect ratio. This is not the case for conventional high temperature annealing, in which nanostructures irreparably acquire an ill-defined blubbery geometry. Nanostructures fabricated with this technique retain their designed optical properties, as demonstrated with different systems, including Fano-resonant structures and metasurfaces built from a collection of similar nanostructures with different stoichiometries. The ellipsometric data of those alloys agree well with theoretical models. The successful fabrication of meta-holograms and meta-lenses demonstrates that this new fabrication technique unlocks additional degrees of freedom for the realization of advanced optical elements at the nanoscale.

## Acknowledgements

DR acknowledges the Swiss Government Excellence scholarship. Funding from the Swiss National Science Foundation (project 200021_162453) and from the European Research Council (ERC-2015-AdG-695206 Nanofactory) is gratefully acknowledged. We thank the staff of the EPFL MicroNanotechnology platform (CMi) for their support and Pierre Mettrraux for XPS measurements.

## Author contributions

DR and CS developed the low temperature annealing process. DR fabricated all the nanostructures reported therein. H-CW designed and measured the metasurfaces. JK measured the spectral response of the nanostructures. All authors analysed the results and contributed to the manuscript.

## References

1   Fichtner, M. Nanotechnological Aspects in Materials for Hydrogen Storage. *Adv. Eng. Mater.* **7**, 443-455, (2005).
2   Lu, J. *et al.* The role of nanotechnology in the development of battery materials for electric vehicles. *Nature Nanotechnology* **11**, 1031-1038, (2016).
3   Hartings, M. R. & Ahmed, Z. Chemistry from 3D printed objects. *Nature Reviews Chemistry* **3**, 305-314, (2019).

4   Burgener, S. *et al.* A roadmap towards integrated catalytic systems of the future. *Nature Catalysis* **3**, 186-192, (2020).
5   Goodnick, S. *et al.* Semiconductor nanotechnology: novel materials and devices for electronics, photonics and renewable energy applications. *Nanotechnology* **21**, 130201, (2010).
6   Wang, Q. H. *et al.* Electronics and optoelectronics of two-dimensional transition metal dichalcogenides. *Nature Nanotechnology* **7**, 699-712, (2012).
7   Liu, C. *et al.* Two-dimensional materials for next-generation computing technologies. *Nature Nanotechnology* **15**, 545-557, (2020).
8   Engel, E. *et al.* Nanotechnology in regenerative medicine: the materials side. *Trends Biotechnol.* **26**, 39-47, (2008).
9   Song, G. *et al.* Emerging Nanotechnology and Advanced Materials for Cancer Radiation Therapy. *Adv. Mater.* **29**, 1700996, (2017).
10  Weiss, J., Takhistov, P. & McClements, D. J. Functional Materials in Food Nanotechnology. *J. Food Sci.* **71**, R107-R116, (2006).
11  Barnes, W. L., Dereux, A. & Ebbesen, T. W. Surface plasmon subwavelength optics. *Nature* **424**, 824, (2003).
12  Thyagarajan, K., Santschi, C., Langlet, P. & Martin, O. J. F. Highly Improved Fabrication of Ag and Al Nanostructures for UV and Nonlinear Plasmonics. *Advanced Optical Materials* **4**, 871-876, (2016).
13  Kottmann, J. P. & Martin, O. J. F. Influence of the cross section and the permittivity on the plasmon-resonance spectrum of silver nanowires. *Appl. Phys. B* **B73**, 299-304, (2001).
14  West, P. R. *et al.* Searching for better plasmonic materials. *Laser & Photonics Reviews* **4**, 795-808, (2010).
15  Naik, G. V., Shalaev, V. M. & Boltasseva, A. Alternative Plasmonic Materials: Beyond Gold and Silver. *Adv. Mater.* **25**, 3264-3294, (2013).
16  Shim, H., Kuang, Z. & Miller, O. D. Optical materials for maximal nanophotonic response. *Optical Materials Express* **10**, 1561-1585, (2020).
17  Zhang, J., Tang, Y., Lee, K. & Ouyang, M. Tailoring light–matter–spin interactions in colloidal hetero-nanostructures. *Nature* **466**, 91-95, (2010).
18  Blaber, M. G., Arnold, M. D. & Ford, M. J. A review of the optical properties of alloys and intermetallics for plasmonics. *Journal of Physics: Condensed Matter* **22**, 143201, (2010).
19  Yang, Y., Zhang, Q., Fu, Z.-W. & Qin, D. Transformation of Ag Nanocubes into Ag–Au Hollow Nanostructures with Enriched Ag Contents to Improve SERS Activity and Chemical Stability. *ACS Applied Materials & Interfaces* **6**, 3750-3757, (2014).
20  Feng, J., Gao, C. & Yin, Y. Stabilization of noble metal nanostructures for catalysis and sensing. *Nanoscale* **10**, 20492-20504, (2018).
21  Gheorghe, D. E. *et al.* Gold-silver alloy nanoshells: a new candidate for nanotherapeutics and diagnostics. *Nanoscale research letters* **6**, 554-554, (2011).
22  Gao, C. *et al.* Fully Alloyed Ag/Au Nanospheres: Combining the Plasmonic Property of Ag with the Stability of Au. *Journal of the American Chemical Society* **136**, 7474-7479, (2014).
23  Gong, C. *et al.* Band Structure Engineering by Alloying for Photonics. *Advanced Optical Materials* **6**, 1800218, (2018).
24  Park, M., Hwang, C. S. H. & Jeong, K.-H. Nanoplasmonic Alloy of Au/Ag Nanocomposites on Paper Substrate for Biosensing Applications. *ACS Applied Materials & Interfaces* **10**, 290-295, (2018).

25  Sanguansap, Y., Karn-orachai, K. & Laocharoensuk, R. Tailor-made porous striped gold-silver nanowires for surface enhanced Raman scattering based trace detection of β-hydroxybutyric acid. *Applied Surface Science* **500**, 144049, (2020).

26  Kahraman, M., Mullen, E. R., Korkmaz, A. & Wachsmann-Hogiu, S. Fundamentals and applications of SERS-based bioanalytical sensing. *Nanophotonics* **6**, 831-852, (2017).

27  Patskovsky, S. *et al.* Hyperspectral reflected light microscopy of plasmonic Au/Ag alloy nanoparticles incubated as multiplex chromatic biomarkers with cancer cells. *Analyst* **139**, 5247-5253, (2014).

28  Tseng, S.-Y. *et al.* Food Quality Monitor: Paper-Based Plasmonic Sensors Prepared Through Reversal Nanoimprinting for Rapid Detection of Biogenic Amine Odorants. *ACS Applied Materials & Interfaces* **9**, 17306-17316, (2017).

29  Wang, P. *et al.* Intracellular and in Vivo Cyanide Mapping via Surface Plasmon Spectroscopy of Single Au–Ag Nanoboxes. *Analytical Chemistry* **89**, 2583-2591, (2017).

30  Qiu, G., Ng, S. P. & Wu, C.-M. L. Bimetallic Au-Ag alloy nanoislands for highly sensitive localized surface plasmon resonance biosensing. *Sensors Actuators B: Chem.* **265**, 459-467, (2018).

31  Chattopadhyay, S., Bysakh, S., Mishra, P. M. & De, G. In Situ Synthesis of Mesoporous TiO2 Nanofibers Surface-Decorated with AuAg Alloy Nanoparticles Anchored by Heterojunction Exhibiting Enhanced Solar Active Photocatalysis. *Langmuir* **35**, 14364-14375, (2019).

32  Kavitha, R. & Kumar, S. G. Review on bimetallic-deposited TiO2: preparation methods, charge carrier transfer pathways and photocatalytic applications. *Chemical Papers* **74**, 717-756, (2020).

33  An, H. *et al.* Design of AgxAu1−x alloy/ZnIn2S4 system with tunable spectral response and Schottky barrier height for visible-light-driven hydrogen evolution. *Chem. Eng. J.* **382**, 122953, (2020).

34  Hwang, C. S. H. *et al.* Ag/Au Alloyed Nanoislands for Wafer-Level Plasmonic Color Filter Arrays. *Scientific Reports* **9**, 9082, (2019).

35  Valenti, M. *et al.* Hot Carrier Generation and Extraction of Plasmonic Alloy Nanoparticles. *ACS Photonics* **4**, 1146-1152, (2017).

36  Davey, W. P. Precision Measurements of the Lattice Constants of Twelve Common Metals. *Physical Review* **25**, 753-761, (1925).

37  Guisbiers, G. *et al.* Electrum, the Gold–Silver Alloy, from the Bulk Scale to the Nanoscale: Synthesis, Properties, and Segregation Rules. *ACS Nano* **10**, 188-198, (2016).

38  Lee, I., Han, S. W. & Kim, K. Production of Au–Ag alloy nanoparticles by laser ablation of bulk alloys. *Chemical Communications*, 1782-1783, (2001).

39  Ferrando, R., Jellinek, J. & Johnston, R. L. Nanoalloys: From Theory to Applications of Alloy Clusters and Nanoparticles. *Chem. Rev.* **108**, 845-910, (2008).

40  Gilroy, K. D. *et al.* Bimetallic Nanocrystals: Syntheses, Properties, and Applications. *Chem. Rev.* **116**, 10414-10472, (2016).

41  Beyene, H. T. *et al.* Preparation and plasmonic properties of polymer-based composites containing Ag–Au alloy nanoparticles produced by vapor phase co-deposition. *JMatS* **45**, 5865-5871, (2010).

42  Gong, C. *et al.* Near-Field Optical Properties of Fully Alloyed Noble Metal Nanoparticles. *Advanced Optical Materials* **5**, 1600568, (2017).

43  Albrecht, W. *et al.* Fully alloyed metal nanorods with highly tunable properties. *Nanoscale* **9**, 2845-2851, (2017).

44 Liu, K. *et al.* Porous Au–Ag Nanospheres with High-Density and Highly Accessible Hotspots for SERS Analysis. *Nano Letters* **16**, 3675-3681, (2016).
45 Zhang, T. *et al.* Controlled synthesis of sponge-like porous Au–Ag alloy nanocubes for surface-enhanced Raman scattering properties. *Journal of Materials Chemistry C* **5**, 11039-11045, (2017).
46 Shore, M. S. *et al.* Synthesis of Au(Core)/Ag(Shell) Nanoparticles and their Conversion to AuAg Alloy Nanoparticles. *Small* **7**, 230-234, (2011).
47 Chen, Y. Nanofabrication by electron beam lithography and its applications: A review. *Microelectronic Engineering* **135**, 57-72, (2015).
48 Nishijima, Y. & Akiyama, S. Unusual optical properties of the Au/Ag alloy at the matching mole fraction. *Optical Materials Express* **2**, 1226-1235, (2012).
49 Xu, L., Tan, L. S. & Hong, M. H. Tuning of localized surface plasmon resonance of well-ordered Ag/Au bimetallic nanodot arrays by laser interference lithography and thermal annealing. *Applied Optics* **50**, (2011).
50 Bai, Y., Gao, C. & Yin, Y. Fully alloyed Ag/Au nanorods with tunable surface plasmon resonance and high chemical stability. *Nanoscale* **9**, 14875-14880, (2017).
51 Wang, X., Santschi, C. & Martin, O. J. F. Strong Improvement of Long-Term Chemical and Thermal Stability of Plasmonic Silver Nanoantennas and Films. *Small* **13**, 1700044, (2017).
52 Lapujoulade, J. The roughening of metal surfaces. *Surface Science Reports* **20**, 195-249, (1994).
53 Kadkhodazadeh, S. *et al.* Optical Property–Composition Correlation in Noble Metal Alloy Nanoparticles Studied with EELS. *ACS Photonics* **6**, 779-786, (2019).
54 Huang, J.-S. *et al.* Atomically flat single-crystalline gold nanostructures for plasmonic nanocircuitry. *Nature Communications* **1**, 150, (2010).
55 Shen, P.-T. *et al.* Temperature- and roughness-dependent permittivity of annealed/unannealed gold films. *Optics Express* **24**, 19254-19263, (2016).
56 Rioux, D. *et al.* An Analytic Model for the Dielectric Function of Au, Ag, and their Alloys. *Advanced Optical Materials* **2**, 176-182, (2014).
57 Peña-Rodríguez, O. *et al.* Optical properties of Au-Ag alloys: An ellipsometric study. *Optical Materials Express* **4**, 403-410, (2014).
58 Argyle, M. D. & Bartholomew, C. H. Heterogeneous Catalyst Deactivation and Regeneration: A Review. *Catalysts* **5**, 145-269, (2015).
59 Kottmann, J. P., Martin, O. J. F., Smith, D. R. & Schultz, S. Dramatic localized electromagnetic enhancement in plasmon resonant nanowires. *CPL* **341**, 1-6, (2001).
60 Lovera, A., Gallinet, B., Nordlander, P. & Martin, O. J. F. Mechanisms of fano resonances in coupled plasmonic systems. *ACS Nano* **7**, 4527-4536, (2013).
61 Kern, A. M. & Martin, O. J. F. Surface integral formulation for 3D simulations of plasmonic and high permittivity nanostructures. *J. Opt. Soc. Am. A* **26**, 732-740, (2009).
62 Gallinet, B., Kern, A. M. & Martin, O. J. F. Accurate and versatile modeling of electromagnetic scattering on periodic nanostructures with a surface integral approach. *J. Opt. Soc. Am. A* **27**, 2261-2271, (2010).
63 Wang, H.-C. *et al.* Ultrathin Planar Cavity Metasurfaces. *Small* **14**, 1703920, (2018).
64 Gramotnev, D. K. & Bozhevolnyi, S. I. Plasmonics beyond the diffraction limit. *Nature Photonics* **4**, 83-91, (2010).
65 Lévêque, G. & Martin, O. J. F. Optical interactions in a plasmonic particle coupled to a metallic film. *Optics Express* **14**, 9971-9981, (2006).
66 Ding, F., Pors, A. & Bozhevolnyi, S. I. Gradient metasurfaces: a review of fundamentals and applications. *Rep. Prog. Phys.* **81**, 026401, (2017).

67  Gao, L., Lemarchand, F. & Lequime, M. Refractive index determination of SiO2 layer in the UV/Vis/NIR range: spectrophotometric reverse engineering on single and bi-layer designs. *Journal of the European Optical Society - Rapid publications; Vol 8 (2013)*, (2013).
68  Ho, Y. Z. *et al.* Anomalous reflection from metasurfaces with gradient phase distribution below $2\pi$. *Applied Physics Express* **9**, 072502, (2016).
69  Huang, Y.-W. *et al.* Aluminum Plasmonic Multicolor Meta-Hologram. *Nano Lett.* **15**, 3122-3127, (2015).
70  Wang, S. *et al.* Broadband achromatic optical metasurface devices. *Nature Communications* **8**, 187, (2017).

# Supporting information

# Low temperature annealing method for fabricating alloy nanostructures and metasurfaces: Unlocking a novel degree of freedom

Debdatta Ray[‡], Hsiang-Chu Wang[‡], Jeonghyeon Kim, Christian Santschi, Olivier J.F Martin[*,]

Nanophotonics and Metrology Laboratory, Swiss Federal Institute of Technology, Lausanne (EPFL), 1015 Lausanne, Switzerland

[‡]These authors contributed equally to this work

[*]corresponding author: olivier.martin@epfl.ch

## XPS measurements

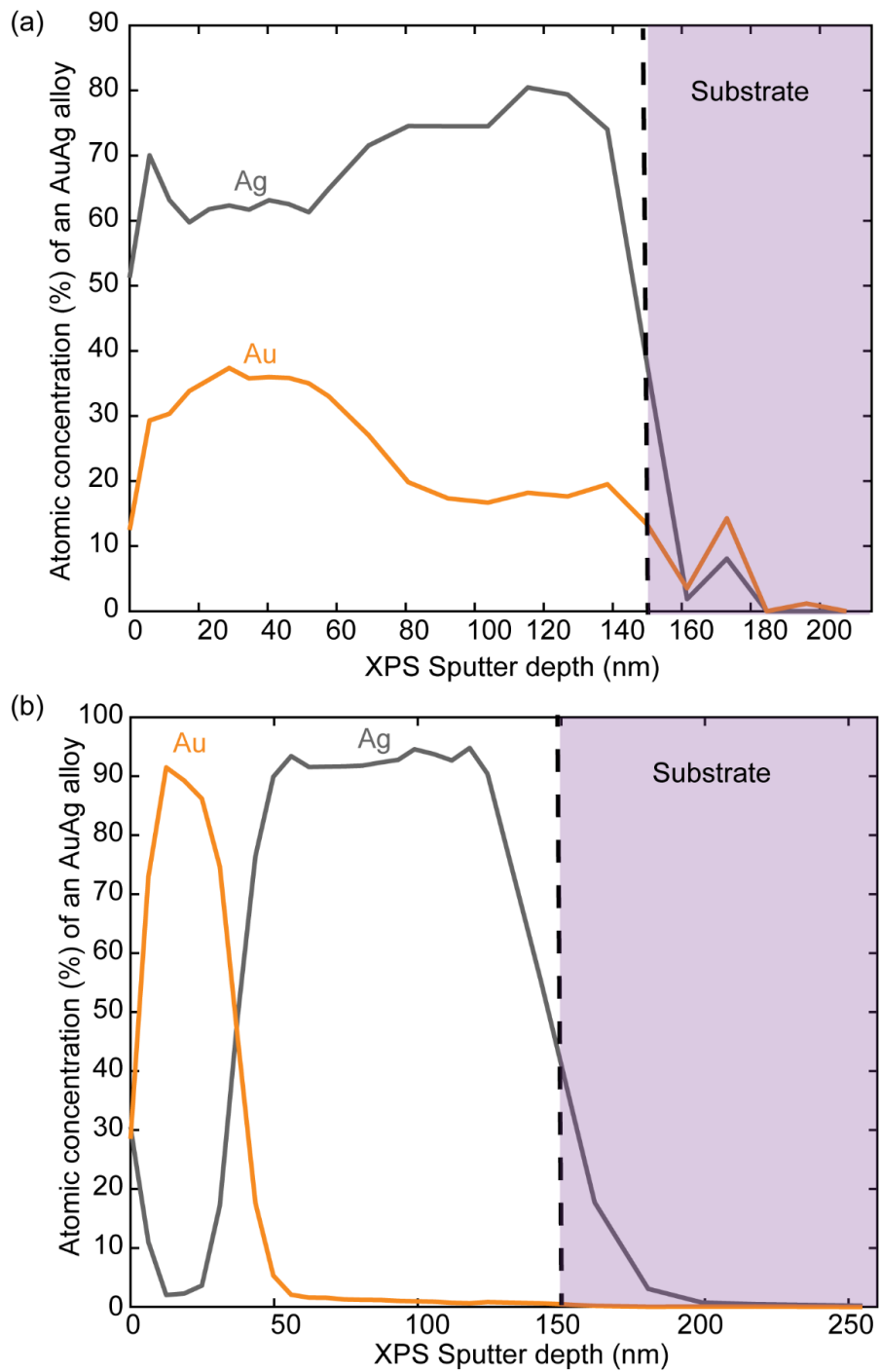

**Fig. SI 1 | a** XPS measurements for an $Au_{0.2}Ag_{0.8}$ films annealed for 300°C for 8 hours. **b** XPS measurements for an $Au_{0.2}Ag_{0.8}$ films not annealed.

## Ellipsometry measurements

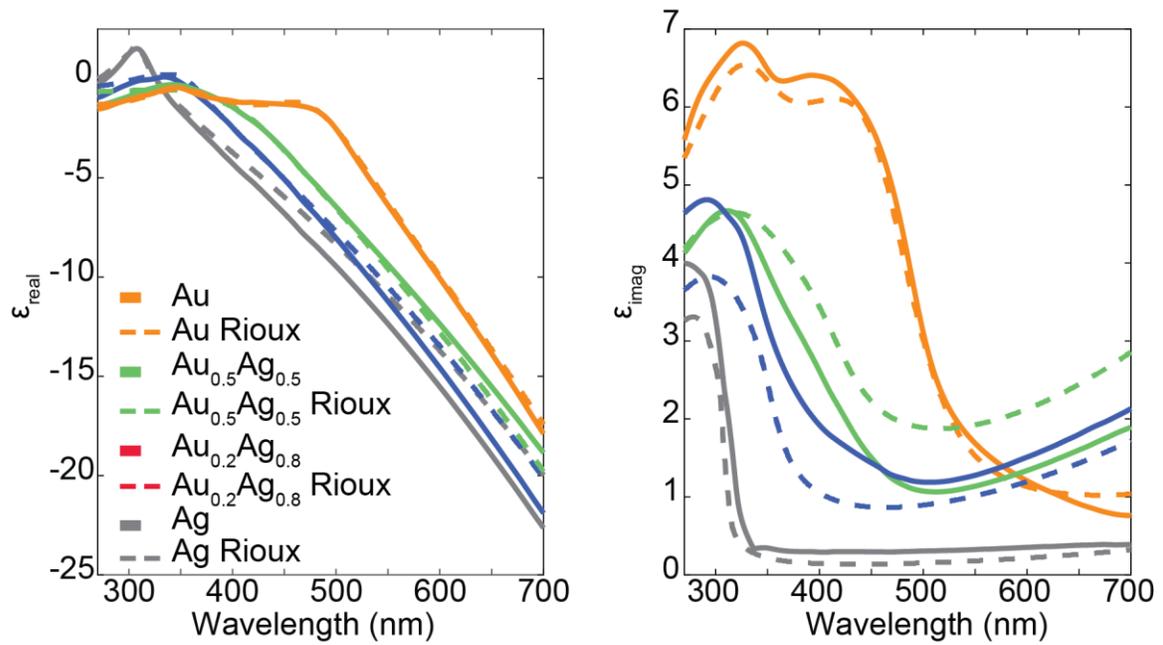

**Fig. SI 2 |** Comparison of permittivity values obtained from the Rioux model and those measured using ellipsometry.

## Ageing of bi-layer films

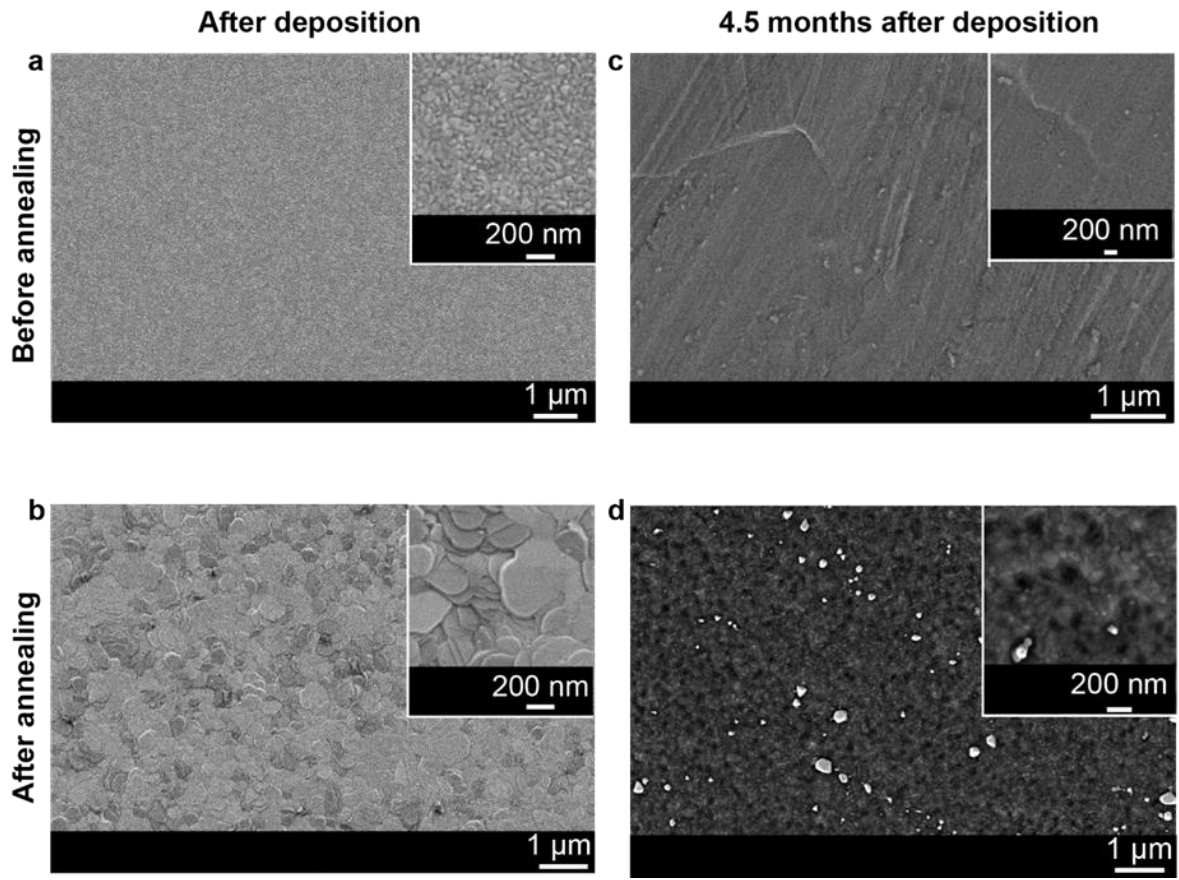

**Fig. SI 3 |** SEM images of a 150 nm thick $Au_{0.8}Ag_{0.2}$ thin films **a** not annealed after deposition, and **b** annealed at 300°C for 8 hours followed by 450°C for 30 min. SEM images of a 150 nm thick $Au_{0.8}Ag_{0.2}$ thin film after 4.5 months **c** not annealed, and **d** annealed at 300°C for 8 hours followed by 450°C for 60 min.

## EDX measurement of nanostructures

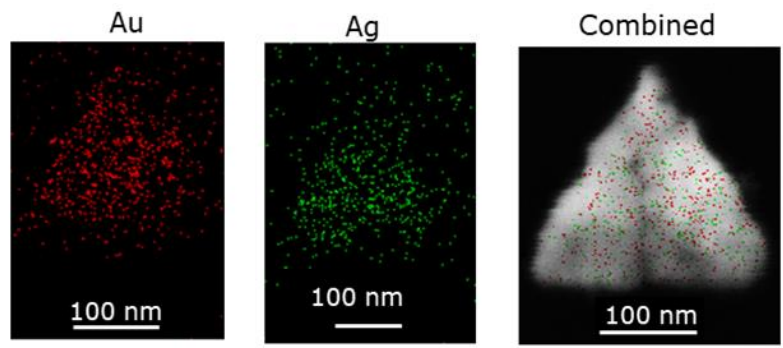

**Fig. SI 4 |** EDX images of the nanotriangle showing distribution of Au, Ag and superimposed image of Au and Ag distributions.

## Darkfield measurement setup

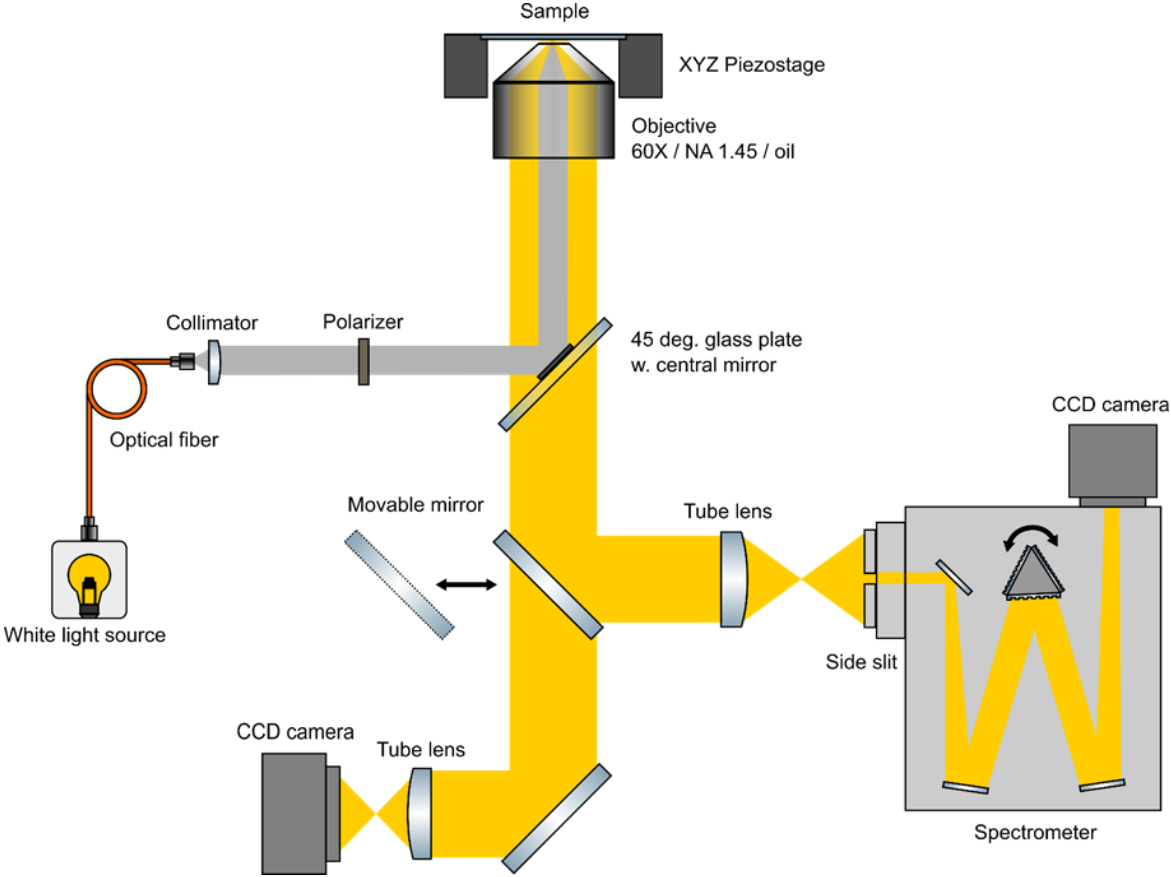

**Fig. SI 5 |** Dark field measurement setup for measuring spectra of Fano resonant strcutures.

## Reflectance measurement setup

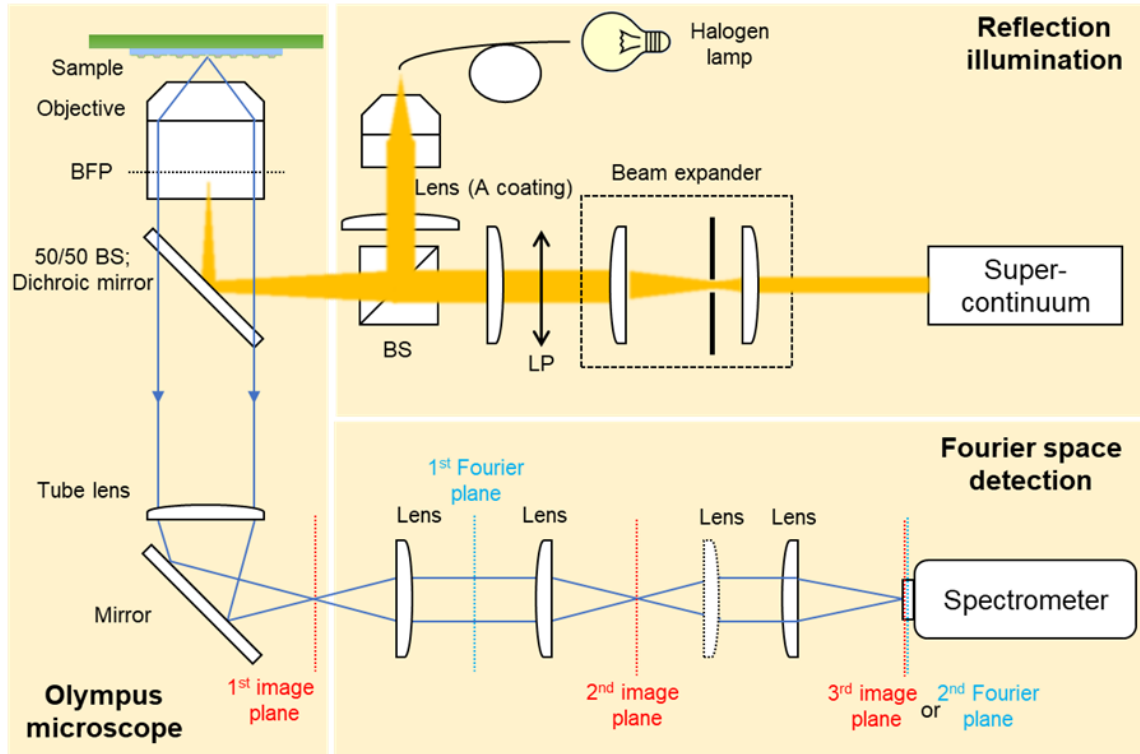

**Fig. SI 6 |** A silver (Ag) mirror is utilized as reference. The sample is illuminated with a Fianium supercontinuum white light source. A linear polarizer is also applied. Spectra are measured with the spectrometer at the 3rd image plane.

## 3-step EBL process for binary metasurface fabrication

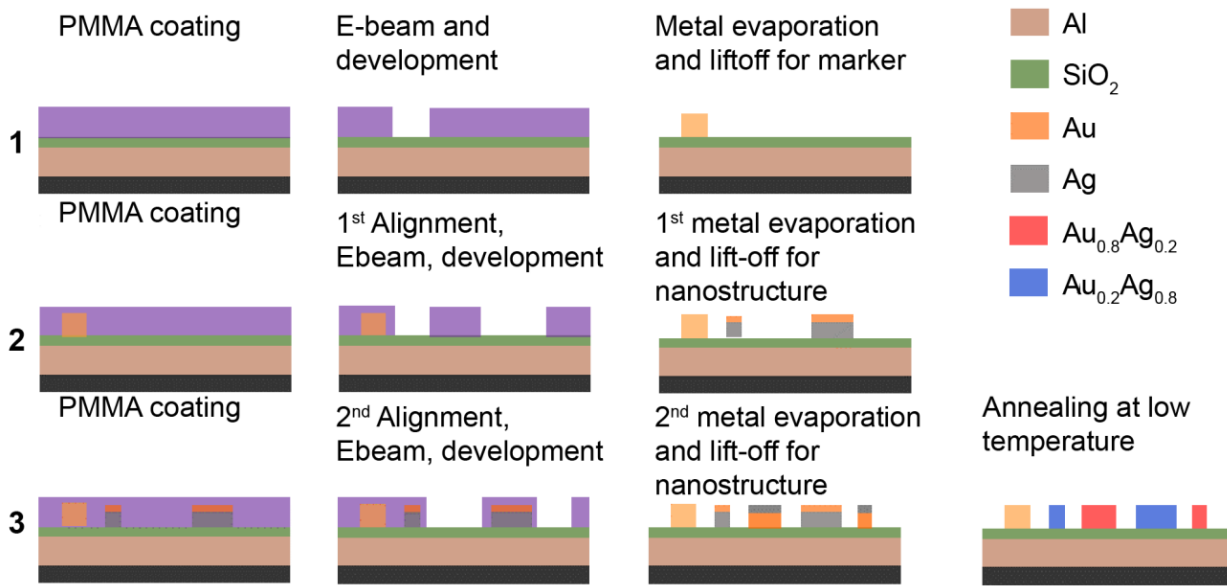

**Fig. SI 7 |** The fabrication of binary metasurfaces built from alloyed nanostructures with two different stoichiometries requires multiple e-beam exposures. The schematic here shows the process flow of the fabrication. The numbers in bold in the left denotes the exposure number.

## Meta-hologram measurement setup

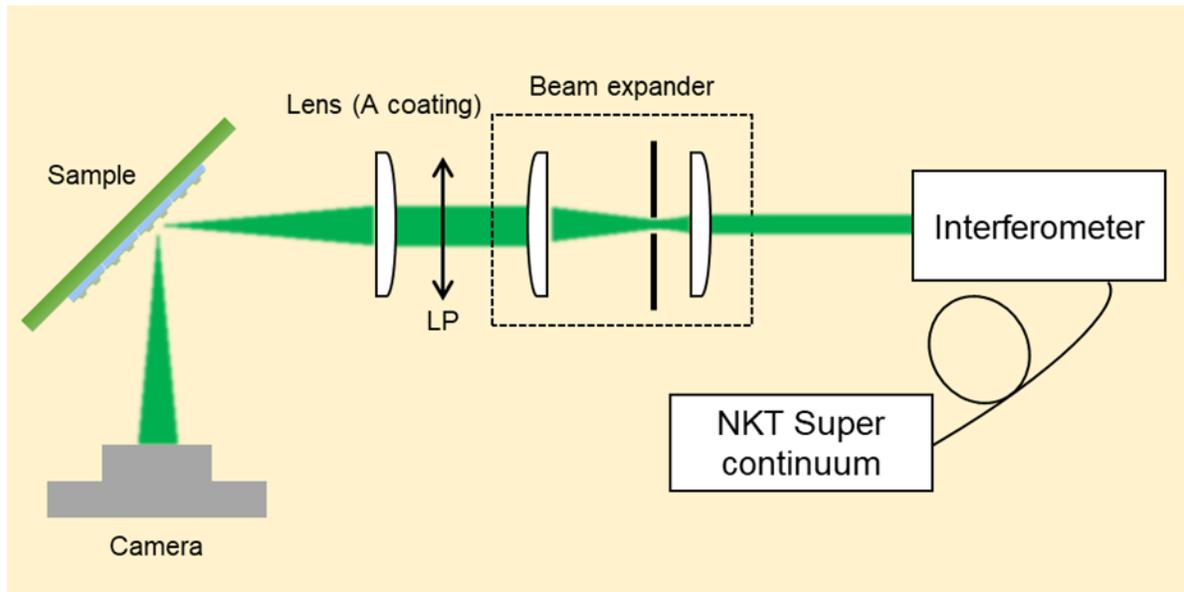

**Fig. SI 8 |** The focal spot is captured at the 2nd Fourier plane. The focal spot can be distinguished from the normal reflection with the sample tilt few degrees. The normal reflection is blocked in the 1st Fourier plane.

## Metalens measurement setup

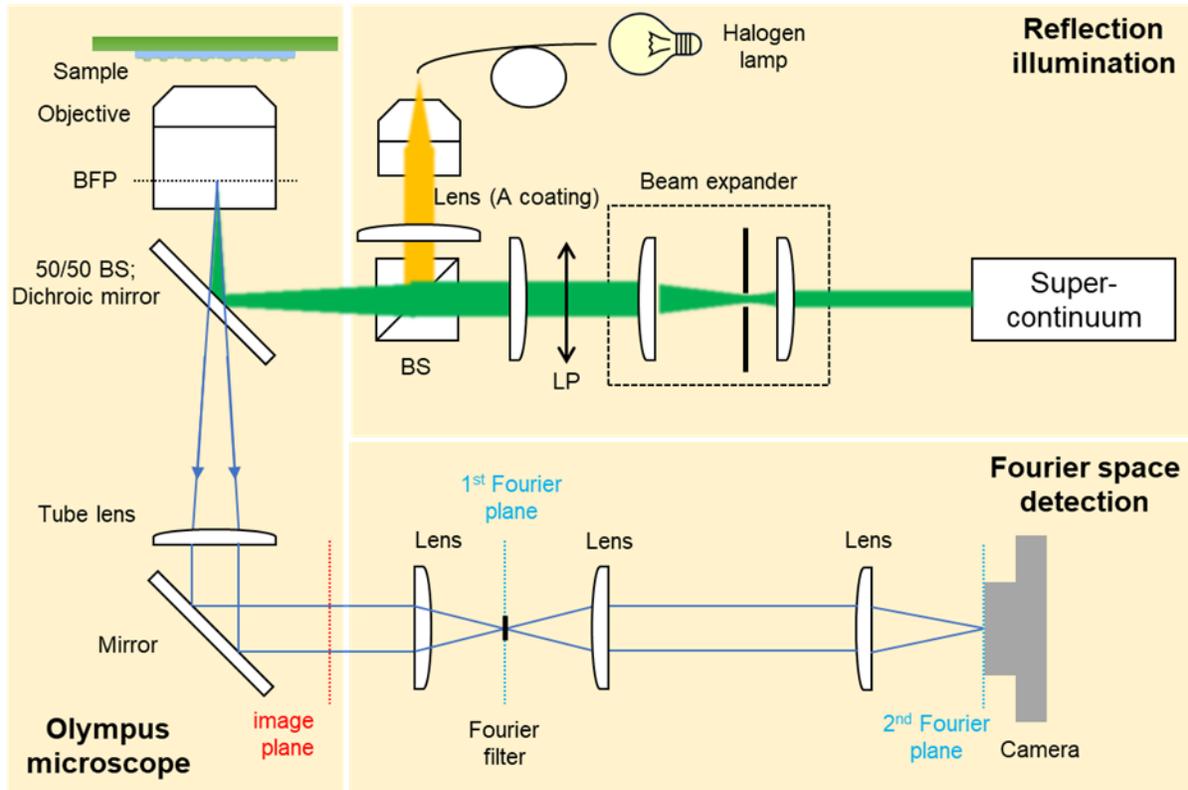

**Fig. SI 9 |** The focal spot is captured at the 2nd Fourier plane. The focal spot can be distinguished from the normal reflection with the sample tilt few degrees. The normal reflection is blocked in the 1st Fourier plane.